# $^{57}$Fe and $^{151}$Eu Mössbauer studies of 3*d*-4*f* spin interplay in $EuFe_{2-x}Ni_xAs_2$

K. Komędera[1], J. Gatlik[1], A. Błachowski[1*], J. Żukrowski[2], D. Rybicki[3], A. Delekta[4], M. Babij[5], and Z. Bukowski[5]

[1]*Mössbauer Spectroscopy Laboratory, Institute of Physics, Pedagogical University of Krakow, ul. Podchorążych 2, 30-084 Kraków, Poland*
[2]*Academic Centre for Materials and Nanotechnology, AGH University of Science and Technology, Av. A. Mickiewicza 30, 30-059 Kraków, Poland*
[3]*AGH University of Science and Technology, Faculty of Physics and Applied Computer Science, Av. A. Mickiewicza 30, 30-059 Kraków, Poland*
[4]*Institute of Geography, Pedagogical University of Krakow, ul. Podchorążych 2, 30-084 Kraków, Poland*
[5]*Institute of Low Temperature and Structure Research, Polish Academy of Sciences, ul. Okólna 2, 50-422 Wrocław, Poland*

[*]Corresponding author: sfblacho@cyf-kr.edu.pl

## Abstract

The $EuFe_{2-x}Ni_xAs_2$ (with $0 \le x \le 0.4$) compounds exhibiting 3*d* and/or 4*f* magnetic order were investigated by means of $^{57}$Fe and $^{151}$Eu Mössbauer spectroscopy. Additionally, results for $EuNi_2As_2$ are reported for comparison. It was found that spin-density-wave order of the Fe itinerant moments is monotonically suppressed by Ni-substitution. However, the 3*d* magnetic order survives at the lowest temperature up to at least $x = 0.12$ and it is certainly completely suppressed for $x = 0.20$. The Eu localized moments order regardless of the Ni concentration, but undergo a spin reorientation with increasing x from alignment parallel to the *a*-axis in the parent compound, toward *c*-axis alignment for $x > 0.07$. Change of the 4*f* spins ordering from antiferromagnetic to ferromagnetic takes place simultaneously with a disappearance of the 3*d* spins order what is the evidence of a strong coupling between magnetism of $Eu^{2+}$ ions and the conduction electrons of $[Fe_{2-x}Ni_xAs_2]^{2-}$ layers. The Fe nuclei experience the transferred hyperfine magnetic field due to the $Eu^{2+}$ ordering for Ni-substituted samples with $x > 0.04$, while the transferred field is undetectable in $EuFe_2As_2$ and for compound with a low Ni-substitution level. It seems that the 4*f* ferromagnetic component arising from a tilt of the $Eu^{2+}$ moments to the crystallographic *c*-axis leads to the transferred magnetic field at the Fe atoms. Superconductivity is not observed down to 1.8 K, although a comparison with $^{57}$Fe and $^{151}$Eu Mössbauer data for $EuFe_2As_2$-based superconductors indicates a similar magnetic structure.

## 1. Introduction

The $EuFe_2As_2$-based compounds are unique laboratory for investigations of interplay between magnetism and superconductivity (SC), as well as they are a playground for peculiar competition between itinerant 3*d* magnetic order of the spin-density-wave (SDW) type and the localized 4*f* magnetic moments. $EuFe_2As_2$ belongs to the "122" family of parent compounds of iron-based superconductors, the same as $AFe_2As_2$ (A = Ca, Sr, Ba) [1]. It crystallizes in the tetragonal $ThCr_2Si_2$-type structure with a space group *I*4/*mmm* at room temperature, distorted into orthorhombic structure with space group *Fmmm* upon cooling. Structural transition is accompanied by the Fe-3*d* itinerant SDW ordering at $T_{SDW} = 190$ K with saturation moment $M_{sat} = 0.99\ \mu_B$ along *a*-axis. The Eu-4*f* localized magnetic moments order in an antiferromagnetic (AFM) A-type structure at $T_N = 19$ K with effective moment $\mu_{eff} = 7.94\ \mu_B$ [2]. The neutron diffraction studies suggest rather weak coupling between the Fe and Eu magnetism, but it was found that the strength of interplay between 3*d* and 4*f* electrons can be tunable by chemical doping. The Fe magnetic order drives the structural phase transition, which indicates a strong coupling between structural and magnetic components [2]. The $Eu^{2+}$ ions are located in the planes perpendicular to the *c*-axis separating $[Fe_2As_2]^{2-}$ layers. They are in a divalent $^8S_{7/2}$ state without orbital contribution to the 4*f* magnetic moments and a spin-only moments are aligned along the crystallographic *a*-axis. The A-type order of $Eu^{2+}$ moments means that they are coupled ferromagnetically within the plane and antiferromagnetically between the planes. Adjacent europium planes are about 6 Å apart [2], hence direct overlap of interplanar 4*f* orbitals can be neglected. It can be assumed that the charge carrier mediated Ruderman-Kittel-Kasuya-Yosida (RKKY) interaction is responsible for the AFM exchange between interplanar Eu moments. An interesting observation is a field-induced spin reorientation in the presence of a magnetic field along both the *a* and *c* axes, which causes that the ground-state AFM configuration of $Eu^{2+}$ moments transforms into a ferromagnetic (FM) structure with moments along the applied field direction [3]. Another interesting effect for $EuFe_2As_2$ is that the SDW transition is suppressed by applied pressure and bulk superconductivity appears in the pressure range from 2.5 to 3 GPa [4, 5]. The $Eu^{2+}$ AFM order and the Néel temperature $T_N$ reveal no changes up to 3 GPa, then the applied pressure above 6 GPa causes a 4*f* spin reorientation and the $Eu^{2+}$ FM order, while further pressure increase above 8 GPa results in suppression of the FM state, which is connected with the Eu valence change from a divalent state to a nearly trivalent state [6, 7].

Superconductivity in $EuFe_2As_2$-based compounds can be achieved by substitutions on either one of three lattice sites as well as by pressure. There are many possibilities for hole and electron doping or isovalent substitution leading to SC, e.g. $Eu_{1-x}A_xFe_2As_2$ with A = K, Na, La, and $EuFe_{2-x}T_xAs_2$ with T = Co, Ru, Ir, or $EuFe_2(As_{1-x}P_x)_2$ [1]. It is interesting to note, that 4*f* magnetic order occurs within superconducting material as well, and doping leading to superconductivity usually causes some 4*f* magnetic moment reorientation (canting) with a generation of the 4*f* FM component [8-10]. The iron-based superconductors are known as materials for which the SC occurs in proximity to a magnetic instability and the magnetic fluctuations could play a key role in the Cooper pairs formation. Hence, it is important to establish the nature of magnetism in these materials.

The electron doping is achieved by partial substitution of the Fe by Ni in $EuFe_{2-x}Ni_xAs_2$ system with structural stability up to x = 0.4, while the Eu sublattice remains chemically intact. Although Ni-substitution causes suppression of the SDW, nevertheless the SC has not been observed down to 2 K [11].

The Mössbauer spectroscopy for $^{57}$Fe and $^{151}$Eu isotopes is a useful local probe for simultaneous investigations of the mutual interaction between the magnetic Fe and Eu sublattices. It has also been shown that the Mössbauer hyperfine parameters are sensitive to change of the electronic charge modulations in the iron-based superconductors [12-14]. A previous $^{57}Fe$ and $^{151}Eu$ Mössbauer investigations have found a significant coupling between the 3*d* and 4*f* magnetic subsystems in the $EuFe_2As_2$-based compounds [8, 10, 15-20]. The transferred magnetic hyperfine field from the 4*f*-Eu order to the Fe nuclei even in the superconducting state was reported for the first time by Nowik and Felner for $EuFe_2(As_{1-x}P_x)_2$ [10]. An enhancement of Fe spin dynamics closely above superconducting critical temperature and related to enhanced Eu spin fluctuations was reported for superconducting $EuFe_2(As_{0.81}P_{0.19})_2$ [15] and $EuFe_2(As_{0.7}P_{0.3})_2$ [16]. $EuFe_{2-x}T_xAs_2$ with the same substitution level $x \approx 0.2$ for T = Co (superconductor) and T = Ni (non-superconductor) show practically the same Mössbauer spectra, confirming that only the Fe magnetic state affects the ordered Eu sublattice direction [17] and the 4*f* electronic system is unaffected by the SC transition. Substitution of Eu site by K atoms gives rise to superconductivity, but long-range 4*f* magnetic order is suppressed due to the magnetic moment dilution at the rare-earth site in $Eu_{0.5}K_{0.5}Fe_2As_2$ system [18]. In $Eu(Fe_{0.86}Ir_{0.14})_2As_2$, the SDW order is fully suppressed and the $Eu^{2+}$ moments order magnetically with an appreciable FM component, which causes resistivity reentrance below the superconducting transition temperature [19]. These brief examples of Mössbauer spectroscopy studies show peculiar phenomena existing in the $EuFe_2As_2$-based systems due to the 3*d*-4*f* spin interplay.

On the other hand, the electron doping by the Ni atoms substituting the Fe atoms causes interesting effects in the iron-pnictide superconducting compounds. In $(Eu_{0.5}K_{0.5})Fe_{2-x}Ni_xAs_2$, the compensation of holes (K doping at the Eu site) by the electrons (Ni doping at the Fe site) leads to the reappearance of the resistivity anomaly presumably due to the SDW transition [21], which is fully suppressed for $(Eu_{0.5}K_{0.5})Fe_2As_2$. In particular, the anomaly associated with the SDW is observed at about 200 K for x = 0.24, which is close to the SDW transition temperature for parent $EuFe_2As_2$. It means that the electron doping almost compensates holes in $(Eu_{0.5}K_{0.5})Fe_{1.76}Ni_{0.24}As_2$ [21]. The itinerant FM ordering associated with the Co/Ni spins was discovered in $Eu(Co_{1-x}Ni_x)_2As_2$ [22], while it is known that the Co atoms do not participate in the magnetic ordering in $EuCo_2As_2$. The same phenomenon occurs in $CaK(Fe_{1-x}Ni_x)_4As_4$ where the itinerant SDW magnetism and superconductivity coexist, while no magnetic hyperfine field (e.g. no static magnetic order) was detected for $CaKFe_4As_4$ superconductor [23]. The Ni doping leads to superconductivity of the "112" compound $EuFeAs_2$, i.e. the magnetism of the 4*f*-Eu order at 44 K and the 3*d*-Fe SDW order at 57 K coexist with the superconducting transition at 14 K for $EuFe_{0.97}Ni_{0.03}As_2$ [24].

The Ni doping of the "122" iron-pnictide parent compounds $AFe_2As_2$ (A = Ca, Sr, Ba) changes these systems in the same following way: 1) the SDW is suppressed with increasing Ni concentration, 2) the SC appears. The maximum critical temperature and optimal substitution level x for $AFe_{2-x}Ni_xAs_2$ (A = Ca [25], Sr [26], Ba [27]) are as follows: 15 K for x = 0.06, 10 K for x = 0.18, and 21 K for x = 0.10, respectively. The same effect of the SDW suppression is observed in the fourth member of "122" family, i.e. $EuFe_{2-x}Ni_xAs_2$ [11]. Hence, it raised the question "why is it not a superconductor?". Of course the suspect culprit is the 4*f* magnetism.

The $EuFe_{2-x}Ni_xAs_2$ system was studied by means of Mössbauer spectroscopy in early years of a fascination with the iron-based superconductors and the results for one composition with x = 0.20 and for selected temperatures were reported [28, 29, 17]. The present contribution reports Mössbauer results obtained by means of $^{57}Fe$ and $^{151}Eu$ spectroscopy applied to systematic investigations of the magnetic properties of the Ni-substituted $EuFe_2As_2$ compounds with an emphasis on study of the 3*d*-4*f* spin interaction.

## 2. Experiment

Single crystals of $EuFe_{2-x}Ni_xAs_2$ were grown using the Sn flux method. The starting materials, high-purity elements of Ca, Fe, Ni, As, and Sn, were taken in the atomic ratio of Eu:Fe:Ni:As:Sn=1:(1-x):10x:2:30 [25]. The constituent components were loaded into alumina crucibles and placed in silica ampules sealed under vacuum. The ampules were heated to 1050°C and kept at this temperature for 10 h to ensure complete dissolving of all components in molten Sn. Next, the ampules were slowly (2–3°C/h) cooled down to 600°C, then the liquid Sn flux was removed by centrifugation. The chemical composition of the obtained crystals was determined using the energy-dispersive x-ray spectroscopy (EDX) and the following Ni content has been received: x = 0, 0.04, 0.07, 0.08, 0.10, 0.12, 0.20, and 0.40. Crystal lattice constants and phase purity were characterized by powder x-ray diffraction (XRD) using PANalytical X'Pert Pro diffractometer. Resistivity measurements were performed by the four probe technique using Quantum Design PPMS system.

$^{57}$Fe Mössbauer spectroscopy measurements were performed in transmission mode for 14.41-keV transition. The samples with x = 0, 0.04, 0.07, 0.10, 0.12, and 0.40 were selected for this investigation. The RENON MsAa-4 spectrometer operated in the round-corner triangular mode and equipped with the LND Kr-filled proportional detector was applied. The He-Ne laser based interferometer was used to calibrate a velocity scale. A single line commercial $^{57}$Co(Rh) source made by RITVERC GmbH was kept at room temperature. The source linewidth $\Gamma_s = 0.106(5)$ mm/s and the effective source recoilless fraction were derived from fit of the Mössbauer spectrum of 10-μm-thick α-Fe foil. The absorbers were prepared using 31 mg of the $EuFe_{2-x}Ni_xAs_2$ samples in the powder form and mixed with a fine powder of $B_4C$ carrier. The absorber thickness (surface density) for $^{57}$Fe Mössbauer measurements amounted to 15 mg/cm$^2$ of investigated material. The SVT-400 cryostat by Janis Research Inc. was used to maintain temperature of absorbers.

$^{151}$Eu Mössbauer transmission spectra for 21.6-keV resonant transition were collected applying $^{151}SmF_3$ source kept at room temperature and a scintillation detector. The samples with x = 0, 0.07, and 0.10 were selected for this investigation. Additionally, the sample of $EuNi_2As_2$ was measured. The absorbers were made in the same way, albeit they contained about twice as much the $EuFe_{2-x}Ni_xAs_2$ material per unit area. Data for $^{57}$Fe and $^{151}$Eu Mössbauer hyperfine parameters were processed by means of the Mosgraf-2009 software within the transmission integral approximation. Europium and iron nuclei in $EuFe_{2-x}Ni_xAs_2$ experience almost axially symmetric electric field gradient (EFG) with the principal component aligned with the *c*-axis of the crystallographic unit cell. The $^{57}$Fe spectra with SDW component were processed by treating the electric quadrupole interaction in the first order approximation. Remaining spectra were processed by application the full Hamiltonian diagonalization in both nuclear states. The europium hyperfine anomaly (the Bohr-Weisskopf effect) was accounted for $^{151}$Eu spectra. The spectral center shifts are reported with respect to the center shift of the room temperature α-Fe or room temperature $^{151}SmF_3$ source, respectively.

### *Mössbauer spectra evaluation within SDW model*

A transmission integral approximation has been applied to fit Mössbauer spectra exhibiting magnetic hyperfine interaction. The SDW magnetism should be viewed as modulations of the spin polarization of the itinerant electrons. The absorption profile of the SDW spectral component was processed by applying a quasi-continuous distribution of the magnetic hyperfine field $B$. For the Fe magnetic moment being collinear with the hyperfine filed, the amplitude of SDW along the direction parallel to the SDW wave vector can be expressed as a series of odd harmonics [30-33]:

$$B(qx) = \sum_{n=1}^{N} h_{2n-1} \sin[(2n-1)qx], \qquad (1)$$

where $B(qx)$, $q$, $x$, and $qx$ denote respectively the magnetic hyperfine field arising from SDW, the wave number of SDW, relative position of Fe atom along the propagation direction of stationary SDW, and phase shift. The symbols $h_{2n-1}$ denote the amplitudes of subsequent harmonics. The index $N$ enumerates maximum relevant harmonic. Amplitudes up to seven subsequent harmonics ($N = 7$) have been fitted to the spectral shape. The constant number of $N$ harmonics was chosen in such a way as to obtain a good-quality fit of the experimental spectrum represented by the parameter $\chi^2$ per degree of freedom of the order of 1.0. The argument $qx$ satisfies the following condition: $0 \le qx \le 2\pi$ due to the periodicity of the SDW. Further details of the Mössbauer spectra evaluation within SDW model can be found in Ref. [30]. Because the average amplitude of SDW described by expression (1) equals zero, thus the root-mean-square amplitude of SDW expressed as $\langle B \rangle = \sqrt{\langle B^2 \rangle} = \sqrt{\frac{1}{2}\sum_{n=1}^{N} h_{2n-1}^2}$ was used as the average magnetic hyperfine field of the SDW [32]. Here, the parameter $\langle B \rangle$ is proportional to the value of the electronic magnetic moment per unit volume.

## 3. Results and discussion

### 3.a Lattice parameters and electrical resistivity

Lattice parameters of the tetragonal unit cell of $EuFe_{2-x}Ni_xAs_2$ at room temperature are shown in **Fig. 1**. In the range from x = 0 to x = 0.40, the lattice constant $a$ increases slightly by about 0.3%, while the constant $c$ contracts by 1.1%. Hence, the volume of the unit cell decreases by 0.5%, leading to the more compact packing of the $[Fe_2As_2]$ layers and a decrease in the distance between Eu planes.

**Fig. 2** shows relative resistivity plotted versus temperature for various nickel concentrations x. A sharp uplift of resistivity is visible for x = 0 and 0.04 at about 190 K and 165 K, respectively, while a broad hump appears for x = 0.07, 0.10 and 0.12 starting at about 130 K, 90 K and 65 K respectively. Both behaviors are caused by the iron magnetism, but with a gradual ordering of 3$d$ spins in the second case. Upon entering the magnetic state, the metallic behavior changes due to partial gapping of the Fermi surface leading to decrease of the carrier concentration. On the other hand, the spin scattering is reduced owing to the increasing spin order. The first mechanism leads to the upturn, while the second to the downturn of the resistivity with lowering of the temperature. The resistivity anomaly associated with SDW order is invisible for x = 0.20 and x = 0.40 due to lack of this type of magnetism for such high nickel concentration. The europium magnetic ordering is seen as much less pronounced kink on the resistivity at about 20 K. The SC is not observed down to 1.8 K for any of these compounds.

### 3.b $^{57}$Fe Mössbauer measurements

#### *Magnetic hyperfine interactions*

$^{57}$Fe Mössbauer spectra measured at selected temperatures of 4.2, 25, 80 and 300 K are shown in **Fig. 3**. The 25 K was chosen because it was the lowest temperature of our measurements without europium magnetic order. Spectra were fitted with the SDW model in

the magnetically ordered region and with a quadrupole doublet otherwise. Apart from the spectra for the parent compound and the spectra for the compound with the highest Ni content, all other spectra were fitted with superposition of two components. Spectra at 300 K have a shape of pseudo-single line due to small electric quadrupole hyperfine interaction. The precise case is that they are unresolved spectral doublets with small average quadrupole splitting ranging from about 0.1 mm/s for the parent compound to about 0.2 mm/s for the compound with the highest substitution level x = 0.40. An asymmetrical spectral broadening is observed with increasing x due to doping-induced disorder and perturbation of the Fe plane site by the Ni atoms. Spectra at 25 K and 80 K have a shape resulting from the coexistence of the magnetic component and the "non-magnetic" second component. The magnetic spectral component associated with SDW order is described by the average magnetic hyperfine field $\langle B \rangle$. The second spectral component is described by the average quadrupole splitting $\langle \Delta \rangle$ and has significantly broadened line-width. The latter component at the low temperatures was fitted applying some hyperfine magnetic field $B_Q$ with a low value of about 1 – 2 T. Further broadening of spectra and change of their shape are observed at 4.2 K and it results from the hyperfine transferred field $B_t$ from ordered magnetic moments of the $Eu^{2+}$ atoms. The presence of two spectral components is somewhat similar to the so-called double-Q magnetic phase with spatially non-uniform magnetization caused by a coherent superposition of two SDWs, leading to both constructive and destructive interference of the SDWs on alternate Fe sites [34]. The spin amplitudes at the Fe sites are nonuniform, vanishing on half of the sites and doubling on the others. But this scenario requires a constant ratio between contribution of both spectral components, which, as will be shown further in the article, is not the case of $EuFe_{2-x}Ni_xAs_2$ system.

Spectra of $EuFe_{1.96}Ni_{0.04}As_2$ compound across magnetic transition are shown in **Fig. 4**. The magnetic component with low hyperfine field of 1.5 T appears at 200 K. Its contribution and the magnetic field increase with decreasing temperature. Half of the iron moments are magnetically ordered at about 175 K and the complete order of all iron spins is observed at 4.2 K with $\langle B \rangle$ = 7.1 T, which is comparable to $\langle B \rangle$ = 8.1 T for the parent compound. The transferred hyperfine field at iron due to the magnetic ordering of europium is undetectable for $EuFe_{2-x}Ni_xAs_2$ with x = 0 and 0.04. Since the Fe plane is in the middle of the distance along the *c*-axis between adjacent Eu planes, hence, in the case of $\theta = 90°$, the $B_t$ is zero due to a cancellation of the internal magnetic induction from the nearest-neighbor Eu planes of FM-aligned moments corresponding to an A-type collinear AFM state. However, it should be mentioned that some small increase of the Fe hyperfine field of about 0.2 T around temperature of the Eu order in the parent compound $EuFe_2As_2$ was reported by Ikeda *et al.* [38]. Processing of our spectra for x = 0.07, 0.10, and 0.12 at 4.2 K gives the transferred field with fairly similar values for both spectral components and with clearly visible difference in the shape of the spectra compared to the temperature of 25 K, i.e. before the Eu magnetism.

The contribution of SDW magnetic component $A$ together with its average magnetic field $\langle B \rangle$ versus temperature are shown in **Fig. 5**. The $EuFe_{2-x}Ni_xAs_2$ compounds with x = 0.07, 0.10, 0.12 contain a predominant contribution of the SDW phase at the lowest measured temperature reaching of about 90%, 80%, 70% respectively. Moreover, a significant reduction of the hyperfine field of the SDW magnetic component to $\langle B \rangle$ = 5.5, 4.8, 4.6 T at 4.2 K was observed, respectively. Temperature evolution of the SDW hyperfine field $\langle B \rangle(T)$ for $EuFe_{1.96}Ni_{0.04}As_2$ was fitted within the model described in Ref. [30] and compared with the results for $EuFe_2As_2$ [30]. The static critical exponent of 0.130(4) and the coherent SDW order temperature of 184(2) K were obtained. We note that the critical exponent in the substituted system $EuFe_{1.96}Ni_{0.04}As_2$ is almost the same as that in the parent compound [30]. The value close to 1/8 indicates that the universality class (1, 2) is retained and it means that the electronic spin system with SDW obeys the Ising model (one-dimensional spin space) and has two dimensions in the configuration space (magnetized planes). It is difficult to draw such conclusions for compounds with x = 0.07, 0.10, 0.12, but one can see that the magnetic ordering starts at about 160, 130 and 100 K, respectively. These temperatures are significantly higher than a temperatures of the resistivity anomaly associated with iron magnetism which are shown in **Fig. 2**. This is because the Mössbauer spectroscopy is much more sensitive to the electron spin density and even to an incoherent spin density wavelets typical for a critical region of the SDW. The low panel of **Fig. 5** shows temperature dependence of a low hyperfine field $B_Q$ resulting from broadening of the second spectral component. Due to the small contribution of this component, the origin of such weak magnetism and even its existence is doubtful.

The average hyperfine field of SDW $\langle B \rangle$ for the $EuFe_2As_2$ at 4.2 K is 8.1 T and this value is related to 0.99 $\mu_B$ magnetic moment determined from the neutron-diffraction measurements [2]. Such small moment results from itinerant character of iron magnetism. The Ni-substitution causes decreasing of the hyperfine field $\langle B \rangle$ and related magnetic moment can be estimated as $\langle B \rangle = \alpha\ \mu_{Fe}$, where $\mu_{Fe}$ is the on-site magnetic moment of iron atom and $\alpha$ denotes the constant, which is specific for a given compound [39]. To convert the hyperfine magnetic field to the iron magnetic moment, we used $\alpha = 8.18$ T/$\mu_B$. The average hyperfine field of SDW at 4.2 K and related 3*d* itinerant magnetic moment of iron versus Ni-substitution x in $EuFe_{2-x}Ni_xAs_2$ system are shown in **Fig. 6**. The magnetic moment of SDW is halved for the Ni substitution x = 0.12 and it decreases linearly in this dopant concentration range. Taking into account the decreasing contribution of SDW component *A* with increasing Ni content, the weighted average field and corresponding weighted average magnetic moment was presented by triangles in **Fig. 6**.

### *Transferred magnetic field from Eu to Fe*

The $EuFe_{1.60}Ni_{0.40}As_2$ displays no magnetic dipole hyperfine interaction in the whole studied temperature range, as the SDW does not develop for such high concentration of nickel. One has sole non-magnetic component with clearly visible influence of the transferred magnetic hyperfine field $B_t = 1.38(2)$ T at 4.2 K from europium order. It gives an opportunity to estimate the angle $\theta$ between the principal component of the electric field gradient (EFG) at the Fe nuclei (i.e. the crystallographic *c*-axis in the present case) and the $Eu^{2+}$ localized magnetic moment. It is assumed that the transferred field points in the same direction as that of the Eu magnetic moment. The angle $\theta = 39(1)°$ is determined reliably for this spectrum and indicates canting of europium moments from *ab*-plane of about $90° - \theta = 51°$. The lack of iron magnetism and almost the same angle $\theta = 36°$ for $B_t = 1.37$ T at 4.2 K was found previously for $EuFe_{1.80}Ni_{0.20}As_2$ by Nowik and Felner [28, 29]. The $Eu^{2+}$ spin reorientation phenomenon will be also discussed later in this article based on data from the $^{151}Eu$ Mössbauer spectroscopy.

It seems that the transferred field of about 1 T is typical for the $EuFe_2As_2$-based compounds with suppressed SDW magnetism and even in the superconducting state. The transferred magnetic hyperfine field acting in the superconducting layer was observed for the first time in the $EuFe_2(As_{1-x}P_x)_2$ system with x > 0.2 [10]. The advantage of this discovery was that the coexistence of SC and magnetic field in the Fe–As based layer was observed on the same sample under the same experimental conditions, by the local probe, i.e. $^{57}Fe$ Mössbauer spectroscopy. The value of $B_t = 0.93(5)$ T was reported for optimally doped superconductor with x = 0.3, while $B_t = 0.85(2)$ T for x = 0.75, and 0.97(2) T for x = 1, i.e. $EuFe_2P_2$ [10]. For the latter case, the angle $\theta = 15(5)°$ was obtained. Another report shows $B_t = 1.2(1)$T and $\theta = 40°$ for the optimally doped $EuFe_2(As_{1-x}P_x)_2$ superconductor with x = 0.28 [15]. Almost the same values of the $B_t$ for different substitution level x indicate that the electron spin density at the position of the Fe nucleus is the same whether the system is normal or superconducting. Another examples of the Eu to Fe magnetic field transfer and the Eu canting are: $Eu(Fe_{0.71}Co_{0.29})_2As_2$ with $B_t = 1.27(2)$T and $\theta = 40(1)°$ [8], or $Eu(Fe_{0.75}Ru_{0.25})_2As_2$ with $B_t = 0.71(2)$ T [35]. An interesting results of the $B_t = 0.6$ T were reported for the ferromagnetic superconductors $AEuFe_4As_4$ (A = Rb [36] and Cs [37]) with Eu moments laying in the *ab*-plane.

### *Electric hyperfine interactions and relative spectral area*

Temperature dependencies of the spectral center shift δ and the quadrupole splitting Δ for both spectral components (if present) are shown in **Fig. 7**. The SDW phase seems to have a lower value of the δ (i.e. a higher value of the *s*-electron charge density) than the "non-magnetic" phase in cases of compounds and temperature ranges where both components occur. But some effect of a fitting artifact can't be excluded due to a low content of one of the components in the temperature ranges discussed. Hence, the weighted average value of spectral center shift should be considered as a reliable parameter. Generally, the δ is almost insensitive to Ni-substitution level x and keeps the value of about 0.41 and 0.54 mm/s at temperature of 300 and 4.2 K, respectively. The same insignificant Ni-concentration dependence of the δ was found for $CaK(Fe_{1-x}Ni_x)_4As_4$ superconductors [23]. It means that Ni-doping does not noticeably affect the *s*-electron charge density at the Fe nuclei. From this Mössbauer spectroscopy point of view, the Ni should be considered as isovalent dopant in the $EuFe_{2-x}Ni_xAs_2$ system. However, other effects leading to such an outcome, such as opposite influence of the lattice volume effect and the *d*-electrons shielding effect, cannot be ruled out. Hence, based on these results, the Ni atoms commonly recognized as electron dopant in the $EuFe_{2-x}Ni_xAs_2$ can't be called into question**.** The value of δ indicates that Fe atoms are in a low-spin Fe(II) electronic configuration.

The temperature evolution of weighted average spectral center shift <δ>, shown in **Fig. 7,** represents a typical second-order Doppler shift dependence on temperature and it could be treated in terms of the Debye model for the lattice vibrations of iron atoms. The Debye temperatures $\Theta_D$ obtained on the basis of this model are typical for a strongly bound metal-covalent system. It increases with Ni substitution level x from about 400 K for the parent compound to about 500 K for x = 0.10. On the other hand, further increasing x to 0.40 causes the return to the almost initial value of $\Theta_D$. It means that the (Fe/Ni)–As bonds become stiffer with increasing Ni content up to x = 0.10. The x = 0.10 is close to the critical substitution level for the AFM-FM transition. It seems that some magneto-elastic effect accompanies this critical Ni concentration resulting in lattice hardening and jump of $\Theta_D$.

The right panel of **Fig. 7** shows the parameter of the electric quadrupole hyperfine interaction, i.e. the quadrupole spitting Δ. At 300 K, the Δ systematically increases from about 0.12 mm/s for the parent compound to about 0.23 mm/s for the Ni-substitution x = 0.40. It means that the EFG at iron site increases with x in $EuFe_{2-x}Ni_xAs_2$ system. This is a symptom of an increase in disorder and perturbation of the Fe atoms surrounding in Fe-plane caused by the Ni atoms. The absolute value of Δ for the "non-magnetic" phase increases as the temperature decreases. At 4.2 K, the Δ of the SDW phase keeps almost constant and negative value of about -0.10 mm/s. This is a manifestation of a greater strength of the magnetic dipole hyperfine interaction than the electric quadrupole interaction, which masks the EFG perturbation caused by increasing Ni substitution. It should be noted that additional magnetic transferred field form europium order perturbs the Mössbauer spectra at 4.2 K. Accordingly, when the 3*d* magnetism of SDW is weakened by Ni-substitution x > 0.04, the 4*f* transferred magnetism causes difference of the Δ between 4.2 and 25 K. For x = 0.40, for which the iron 3$d$ magnetism was not observed in whole studied temperature range, the 4$f$ magnetism at 4.2 K causes a positive sign of the $\Delta$ = +0.224(7) mm/s (point not shown in **Fig. 7**).

The Mössbauer recoilless fraction can be approximated by the relative spectral area (*RSA*), which is defined as:

$$RSA = \left(\frac{1}{C}\right)\sum_{n=1}^{C}\frac{N_0 - N_n}{N_0},$$

where $C$ denotes the number of data channels for folded spectrum, $N_0$ is the average number of counts per channel far-off resonance, namely the baseline, and $N_n$ stands for the number of counts in the channel $n$. The *RSA* can be directly evaluated from measured spectra as a quantity independent of any physical model. The temperature evolutions of the relative spectral areas for $EuFe_{2-x}Ni_xAs_2$ with x = 0.04, 0.10, 0.40 are displayed in **Fig. 8**. Here, the *RSA*s are calculated for the Mössbauer spectra recorded versus increasing temperature in the non-interrupted series of measurements carried out in the same experimental conditions (radioactive source, geometry, velocity scale, number of data channels, background under the resonance line, and linear response range of the detector system) and for the same mass of each of the samples. The *RSA* slightly increases with decreasing temperature due to the thermal effect. Some slight jump of *RSA* at about 180 K for x = 0.04 is observed due to an increase in the recoilless fraction probably arising from the weak magneto-elastic effect, which results in a fine hardening of lattice vibrations upon development of the magnetic order. Such an effect, but much stronger, is observed upon magnetic transition for other compounds with the iron–pnictogen bonds [40, 41]. In this case, it may be suppressed due to the Ni perturbation of the Fe-plane. For comparison, the compound with x = 0.40 without any 3$d$ magnetism has almost flat temperature dependence of *RSA* and without any irregularity.

### 3.c $^{151}$Eu Mössbauer measurements

It should be noted that $^{151}$Eu Mössbauer spectroscopy is very sensitive to the $\theta$ angle between principal component of EFG and hyperfine magnetic field $B$ on the $^{151}$Eu nucleus, i.e. for the angle between the $c$-axis (in this case of tetragonal or orthorhombic structure) and the $Eu^{2+}$ magnetic moment. The spin reorientation phenomenon can be easily recognized (even by eye) due to characteristic change of the relative intensity of spectral lines, see **Fig. 9**.

$^{151}$Eu Mössbauer spectra for the parent compound and the Ni-substituted compounds with x = 0.07 and 0.10 are shown in **Fig. 10**, while obtained hyperfine parameters are listed in **Table 1**. The symbol $\varepsilon = \frac{1}{4}(c/E_0)eQ_gV_{zz}$ stands for the quadrupole coupling constant for $Eu^{2+}$ under assumption that the EFG is axially symmetric with the main axis aligned with the $c$-axis of the unit cell. The symbol $c$ stands for the speed of light in vacuum, the symbol $E_0$ denotes transition energy, while the symbol $e$ stands for the positive elementary charge. The symbol $Q_g$ = +0.903 b denotes spectroscopic nuclear electric quadrupole moment in the ground state of $^{151}$Eu, while the symbol $V_{zz}$ stands for the principal component of EFG on $Eu^{2+}$ nucleus. For spectra with the europium hyperfine magnetic field one obtained the quadrupole shift $\varepsilon = \frac{1}{4}(c/E_0)eQ_gV_{zz}(3\cos^2\theta - 1)/2$.

**Table 1.** $^{151}$Eu Mössbauer spectroscopy parameters for $EuFe_{2-x}Ni_xAs_2$ and $EuNi_2As_2$. The values for 4.2 K and 5.4 K with superscript "a" were obtained for spectra fitting with fixed angle $\theta$, while superscript "b" corresponds to fitting with free angle $\theta$. The values of respective parameter $\chi^2$ are shown for comparison of the spectra fitting quality. Symbol $\Gamma_a$ stands for the absorber linewidth within transmission integral approximation, while the source linewidth $\Gamma_s$ = 0.72 mm/s was kept constant with a value close to the natural width

$\Gamma_0$ = 0.655 mm/s. The meaning of the other symbols is described in the text. The **Fig. 10** and **Fig. 11** show spectra for $4.2^b$ K and $5.4^b$ K, respectively. Previous results from Ref. [17] and [42] are shown for comparison. Note that the angle $\theta$ is undefined in case of the $V_{zz}$ or $B$ is equal zero.

| $EuFe_{2-x}Ni_xAs_2$ | $T$ (K) | δ (mm/s) | $V_{zz}$ ($10^{22}$V/m$^2$) | ε (mm/s) | $B$ (T) | $\theta$ (°) | $\Gamma_a$ (mm/s) | $\chi^2$ |
|---|---|---|---|---|---|---|---|---|
| x = 0 | 300 | -11.46(1) | -0.39(2) | -1.55 | - | - | 1.52(4) | |
| | $4.2^a$ | -11.40(1) | -0.49(2) | +0.96 | 26.59(3) | 90 | 1.38(2) | 0.44 |
| | $4.2^b$ | -11.40(1) | -0.56(4) | +0.95 | 26.62(4) | 77(3) | 1.36(2) | 0.44 |
| x = 0.07 | 300 | -11.59(1) | -0.37(2) | -1.45 | - | - | 1.58(4) | |
| | $4.2^a$ | -11.79(2) | -0.20(3) | +0.41 | 26.74(5) | 90 | 1.72(4) | 0.68 |
| | $4.2^b$ | -11.80(2) | -0.20(5) | +0.38 | 26.74(6) | 83(6) | 1.74(4) | 0.68 |
| x = 0.10 | 300 | -11.43(1) | -0.34(1) | -1.36 | - | - | 1.42(4) | |
| | $4.2^a$ | -11.43(2) | -0.35(1) | -1.40 | 27.37(4) | 0 | 1.52(4) | 0.53 |
| | $4.2^b$ | -11.44(2) | -0.56(4) | -1.32 | 27.53(5) | 31(2) | 1.54(4) | 0.52 |
| x = 0.20 [17] | 5 | -11.75(2) | | -2.35 | 28.7(2) | ~ 34 | | |
| $EuNi_2As_2$ | 300 | -10.17(1) | -0.23(2) | -0.92 | - | - | 1.58(4) | |
| | $5.4^a$ | -10.01(2) | -0.28(2) | +0.56 | 32.43(4) | 90 | 1.52(2) | 0.53 |
| | $5.4^b$ | -10.01(2) | -0.58(5) | +0.54 | 32.53(5) | 65(2) | 1.50(2) | 0.48 |
| | 4.2 [42] | -9.9(2) | ~ 0 | ~ 0 | 36.5(5) | - | | |

The hyperfine magnetic field $B$ on the $Eu^{2+}$ ions in the spectrum collected at lowest temperatures for the parent $EuFe_2As_2$ and for x = 0.07 are almost the same. The angle $\theta$ equal to or close to the right angle indicates that AFM order of europium 4*f* spins between subsequent *ab*-planes is preserved for these compounds. However, it must be admitted that by fitting the Mössbauer spectra at 4.2 K for $EuFe_2As_2$ and $EuFe_{1.93}Ni_{0.07}As_2$ one obtained a slight canting of the $Eu^{2+}$ moments out of the *ab*-plane of 13(3)° and 7(6)°, respectively. The same result of 13(7)° and ~10° was indicated for $EuFe_2As_2$ studied by means of Mössbauer spectroscopy [10] and magnetic torque measurements [43], respectively. To our knowledge, the canting of $Eu^{2+}$ moments out of the *ab*-plane for $EuFe_2As_2$ has not been observed (yet) by the neutron diffraction and resonant x-ray measurements. Hence, a wary notation $\theta \sim 90°$ is used in **Fig. 10**.

On the other hand, for x = 0.10 the angle $\theta$ between hyperfine field on $Eu^{2+}$ and the main component of EFG on the same ion amounts to 31(2)° at 4.2 K. The main component of EFG is oriented along the crystallographic *c*-axis, while the hyperfine field is aligned with the $Eu^{2+}$ magnetic moment. Hence, the magnetic moment of europium is tilted by $90° - \theta = 59(2)°$ from the *ab*-plane, while in the parent compounds is perpendicular to the *c*-axis. It means that in $EuFe_{1.90}Ni_{0.10}As_2$ magnetic moments of $Eu^{2+}$ tend to rotate on the *c*-axis in similarity to the superconducting $EuFe_2As_2$-based compounds with some canting leading to the ferromagnetic component of the 4*f* origin [8, 10].

Another proof of the AFM to FM transition in $EuFe_{1.90}Ni_{0.10}As_2$ is the hyperfine field $B$ = 27.5 T, hence about 1 Tesla higher than for the parent compound and the compound with a slightly lower substitution level x = 0.07. Increased hyperfine field for $Eu^{2+}$ spins almost aligned along *c*-axis was observed for $EuFe_2As_2$-based substituted compounds [10].

Both phenomena, i.e. the spin reorientation and the accompanying increase in the hyperfine field, were previously observed by means of the $^{151}Eu$ Mössbauer spectroscopy for other substituted $EuFe_2As_2$-based compounds. For $Eu(Fe_{0.75}Ru_{0.25})_2As_2$ superconductor with $T_{sc}$ = 23 K, the $Eu^{2+}$ spins order ferromagnetically below 19.5 K with the hyperfine field 28.8 T at 5 K tilted away from the *c*-axis by 20(3)° [35], while for $EuRu_2As_2$ one obtained $B$ = 30.2 T and $\theta \sim 0°$ [44]. For $Eu(Fe_{0.86}Ir_{0.14})_2As_2$ superconductor with $T_{sc}$ = 22.5 K and double reentrance at lower temperature, the $Eu^{2+}$ moments order magnetically at 18 K with an appreciable FM component and the Eu hyperfine field of 28.5 T at 4.2 K [19]. The $EuFe_2(As_{1-x}P_x)_2$ system preserves the AFM order for x ≤ 0.2 with the angle $\theta = 77(7)°$ and the hyperfine field $B = 26.2$ T [10, 15]. While, the canting effect and Eu hyperfine field increase with accompanying the Eu to Fe transferred field (~ 1 T) have been observed for x > 0.2, with reported values of $\theta = 12(8)°$ and $B = 28.4$ T for x = 0.3 [16], or $\theta = 22(3)°$ and $B = 30.8$ T for x = 1, i.e. $EuFe_2P_2$ [10, 45].

It can be additionally noted that the FM order of Eu magnetic moments laying in the *ab*-plane was detected by $^{151}Eu$ Mössbauer measurements for superconducting $AEuFe_4As_4$ (A = Rb [36] and Cs [37]) with the europium hyperfine field of 22.5 T and 22.4 T, respectively. In this case, the sheets of Eu atoms are separated by the $AFe_4As_4$ unit which rules out the possibility of any strong interaction between the Eu planes. In particular, no magnetic coupling is expected to exist between the Eu atoms along the *c*-direction. On the other hand, the AFM order was reported for parent $EuFeAs_2$ and superconductor $EuFeNi_{0.03}As_2$ [24] with $B = 29.4$ T and 29.1 T, respectively.

Traces of $Eu^{3+}$ makes about 2.6% contribution to the spectra of $EuFe_{1.90}Ni_{0.10}As_2$ in similarity to the superconducting and/or overdoped $EuFe_2As_2$-based compounds studied previously by $^{151}Eu$ Mössbauer spectroscopy [8]. The $Eu^{3+}$ component of spectra was usually assigned to a unidentified foreign phase or a result of oxidation. But some change in valence of the $Eu^{2+}$ belonging to the main phase due to local chemical pressure induced by nickel substitution cannot be ruled out [46, 8]. However, in the present case one cannot prove that trivalent europium is in the main phase of the sample.

The magnetic structures resulting from $^{57}Fe$ and $^{151}Eu$ Mössbauer spectra are shown in lower panel of **Fig. 10**. For parent compound, the Fe-SDW itinerant moments align along *a* direction and order antiferromagnetically in both *a* and *c* directions, while the Eu localized moments align along *a* direction and order antiferromagnetically in *c* direction only. The reduction of the Fe moments follows for x = 0.07 without changes of the 3*d* and 4*f* spins order directions. The further lowering of 3*d* moment by almost half for x = 0.10 results in the 4*f* spin reorientation.

### *$EuNi_2As_2$*

Additionally, $^{151}Eu$ Mössbauer spectra for compound with iron fully replaced by nickel, i.e. $EuNi_2As_2$, are shown in **Fig. 11** and the hyperfine parameters are listed in **Table. 1**. We used exactly the same $EuNi_2As_2$ single crystal sample (from the same batch) that was used in Ref. [47], only powdered for our use. This compound possess a tetragonal structure (space group *I4/mmm* with $a = 4.1145$ Å and $c = 10.091$ Å) at room temperature, similar to $EuFe_2As_2$, but without structural phase transition and magnetic ordering of the 3*d* transition metal (i.e. Ni). The neutron diffraction measurements of this sample [47] found that the $Eu^{2+}$ moments form an incommensurate antiferromagnetic spiral-like structure below the Néel temperature of $T_N = 15$ K. They align ferromagnetically in the *ab*-plane and rotate spirally by 165.6° around the *c*-axis from layer to layer [47]. It must be noted, however, that our $^{151}Eu$ Mössbauer investigation indicates some canting of the europium moments out of the *ab*-plane and $\theta = 65(2)°$ was obtained. On the other hand, we note that $^{151}Eu$ Mössbauer spectroscopy is not sensitive to the rotation of the $Eu^{2+}$ moments around the *c*-axis in this case due to axially symmetric EFG in $EuNi_2As_2$ (*c*-axis is the main axis of the EFG) and, therefore, a spiral-like order of $Eu^{2+}$ moments cannot be seen by this method. The $EuNi_2As_2$ was studied by $^{151}Eu$ Mössbauer spectroscopy in the past [42], but unlike our results, negligible quadrupole interaction and some different value of the hyperfine field was reported [42].

## 4. Conclusions

$^{57}$Fe Mössbauer measurements on $EuFe_{2-x}Ni_xAs_2$ show that Ni doping monotonically suppresses the SDW ground state of itinerant magnetic moments and iron diamagnetism is achieved for x > 0.12, i.e. certainly for x = 0.20. While, $^{151}$Eu Mössbauer spectra indicate that $Eu^{2+}$ magnetic moments order regardless of the Ni-substitution level. However, for the parent $EuFe_2As_2$ and for low Ni-substituted compounds up to about x ≤ 0.07, the localized 4*f* spins order perpendicular to the *c*-axis, whereas from about x ≥ 0.10, the magnetic ordering of europium changes from AFM to FM with the angle of deviation from the *c*-axis of about 30°. It seems that the FM order of $Eu^{2+}$ spins causes the transferred hyperfine magnetic field sensed by $^{57}$Fe nuclei. But, it should be noted that some transferred field was also detected for $^{57}$Fe spectrum at the lowest temperature for x = 0.07, hence close to the range border of the europium AFM order. However, due to broad magnetic $^{57}$Fe spectrum for x = 0.07 at 4.2 K, the transferred field may be somewhat doubtful for this compound and a hypothesis that mainly FM component of the Eu order is responsible for the transferred field seems reasonable. Another explanation that can possibly be proposed is some shielding effect, i.e. the disclosure of the transferred field only after a significant weakening of the iron SDW magnetism caused by the Ni dopant. Also note that Ni substitution in $EuFe_{2-x}Ni_xAs_2$ reduces *c* lattice constant of the unit cell (see **Fig. 1**) and shortens the distance between the Eu planes, which can strengthens the coupling between them.

Based on these and previous results, it can be concluded that the 4*f*-Eu magnetism in the $EuFe_2As_2$-based systems is strongly affected by the magnetic behavior of the Fe-As layer. As long as Fe-plane remains unperturbed, for the parent $EuFe_2As_2$ and a low substituted compounds for which the 3*d*-SDW order is present, the Fe magnetic anisotropy pulls the Eu magnetic moment direction to be parallel to the *ab*-plane. On the other hand, a chemical substitution leads to a continuous suppression of the SDW ordering and for the sufficiently substituted compounds (specific to a particular substituent), when the Fe magnetism is strongly suppressed or absent, the direction of the Eu moments tend to turn parallel to the *c*-axis and the FM order is partly or fully achieved. The disappearance of the SDW changes the RKKY interaction and increases interplanar coupling which leads to a FM arrangement of the Eu moments. This rearrangement of the 4*f* spins as the result of the disappearance of the 3*d* spins magnetism proves a strong coupling between the magnetism of $Eu^{2+}$ ions and the conduction electrons of $[Fe_{2-x}Ni_xAs_2]^{2-}$ layers. Additionally, the transferred magnetic hyperfine field with value of about 1 Tesla at 4.2 K from the magnetically ordered Eu sublattice to the Fe atoms is observed for substituted samples, but it is undetected for the parent and a low-substituted samples. This effect may result in competition between magnetism and superconductivity in $EuFe_2As_2$-based superconductors, because the Zeeman effect, which arises due to the FM order, strongly disfavors formation of the Cooper pair. However, as shown for many $EuFe_2As_2$-based compounds, this does not prevent the occurrence of superconductivity [36, 37, 10]. Eventually, the reentrant effect in some cases was observed [19, 20]. Another effect is the anisotropic resistivity when the interplay between Eu-ferromagnetism and Fe-superconductivity causes zero resistance and diamagnetism only when the supercurrent flows within *ab*-planes, but in comparison, the out-of-plane resistivity does not go to zero [35].

When trying to answer the question posed in introduction of this article "why is $EuFe_{2-x}Ni_xAs_2$ not a superconductor?", it must be said that the phenomena seen by the Mössbauer spectroscopy and described above for this system are the same as for many other $EuFe_2As_2$-based superconductors. This indicates a similar magnetic structure regardless of whether the system is normal conducting or superconducting. Hence, the reason for the lack of superconductivity in $EuFe_{2-x}Ni_xAs_2$ (at least above 1.8 K) is an open question.

## Acknowledgments

This work was supported by National Science Centre, Poland (Grant No. 2018/29/N/ST3/00705). M.B. and Z.B. acknowledge financial support from National Science Centre, Poland (Grant No. 2017/25/B/ST3/02868). D.R. acknowledges support by National Science Centre, Poland (Grant No. 2018/30/E/ST3/00377). We thank Professor Czesław Kapusta (AGH University of Science and Technology, Kraków, Poland) for helpful discussions.

## References

[1] S. Zapf and M. Dressel, Rep. Prog. Phys. **80**, 016501 (2017).
[2] Y. Xiao, Y. Su, M. Meven, R. Mittal, C. M. N. Kumar, T. Chatterji, S. Price, J. Persson, N. Kumar, S. K. Dhar, A. Thamizhavel, and Th. Brueckel, Phys. Rev. B **80**, 174424 (2009).
[3] Y. Xiao, Y. Su, W. Schmidt, K. Schmalzl, C. M. N. Kumar, S. Price, T. Chatterji, R. Mittal, L. J. Chang, S. Nandi, N. Kumar, S. K. Dhar, A. Thamizhavel, and Th. Brueckel, Phys. Rev. B **81**, 220406(R) (2010).
[4] T. Terashima, M. Kimata, H. Satsukawa, A. Harada, K. Hazama, S. Uji, H. S. Suzuki, T. Matsumoto, and K. Murata, J. Phys. Soc. Jpn. **78**, 083701 (2009).
[5] N. Kurita, M. Kimata, K. Kodama, A. Harada, M. Tomita, H. S. Suzuki, T. Matsumoto, K. Murata, S. Uji, and T. Terashima, Phys. Rev. B **83**, 214513 (2011).
[6] K. Matsubayashi, K. Munakata, M. Isobe, N. Katayama, K. Ohgushi, Y. Ueda, N. Kawamura, M. Mizumaki, N. Ishimatsu, M. Hedo, I. Umehara, and Y. Uwatoko, Phys. Rev. B **84**, 024502 (2011).
[7] R. S. Kumar, Y. Zhang, A. Thamizhavel, A. Svane, G. Vaitheeswaran, V. Kanchana, Y. Xiao, P. Chow, Ch. Chen, and Y. Zhao, Appl. Phys. Lett. **104**, 042601 (2014).
[8] A. Błachowski, K. Ruebenbauer, J. Żukrowski, Z. Bukowski, K. Rogacki, P. J. W. Moll, and J. Karpinski, Phys. Rev. B **84**, 174503 (2011).
[9] W. T. Jin, Y. Xiao, Z. Bukowski, Y. Su, S. Nandi, A. P. Sazonov, M. Meven, O. Zaharko, S. Demirdis, K. Nemkovski, K. Schmalzl, L. M. Tran, Z. Guguchia, E. Feng, Z. Fu, and Th. Brückel, Phys. Rev. B **94**, 184513 (2016).
[10] I. Nowik, I. Felner, Z. Ren, G. H. Cao, and Z. A. Xu, J. Phys.: Condens. Matter **23**, 065701 (2011).
[11] Z. Ren, X. Lin, Q. Tao, S. Jiang, Z. Zhu, C. Wang, G. H. Cao, and Z. A. Xu, Phys. Rev. B **79**, 094426 (2009).
[12] A. K. Jasek, K. Komędera, A. Błachowski, K. Ruebenbauer, Z. Bukowski, J. G. Storey, and J. Karpinski, J. Alloys Compd. **609**, 150 (2014).
[13] A. K. Jasek, K. Komędera, A. Błachowski, K. Ruebenbauer, H. Lochmajer, N. D. Zhigadlo, and K. Rogacki, J. Alloys Compd. **658**, 520 (2016).
[14] K. Komędera, J. Gatlik, A. Błachowski, J. Żukrowski, T. J. Sato, D. Legut, and U. D. Wdowik, Phys. Rev. B **103**, 024526 (2021).
[15] T. Goltz, S. Kamusella, H. S. Jeevan, P. Gegenwart, H. Luetkens, P. Materne, J. Spehling, R. Sarkar, and H.-H. Klauss, J. Phys.: Conf. Ser. **551**, 012025 (2014).
[16] J. Munevar, H. Micklitz, M. Alzamora, C. Argüello, T. Goko, F. L. Ning, T. Munsie, T. J. Williams, A. A. Aczel, G. M. Luke, G. F. Chen, W. Yu, Y. J. Uemura, and E. Baggio-Saitovitch, Solid State Comm. **187**, 18 (2014).
[17] I. Nowik, I. Felner, Z. Ren, G. H. Cao, and Z. A. Xu, New J. Phys. **13**, 023033 (2011).
[18] Anupam, P.L. Paulose, H. S. Jeevan, C. Geibel, and Z. Hossain, J. Phys.: Condens. Matter **21**, 265701 (2009).

[19] U. B. Paramanik, P. L. Paulose, S. Ramakrishnan, A. K. Nigam, C. Geibel, and Z Hossain, Supercond. Sci. Technol. **27**, 075012 (2014).
[20] K. Komędera, A. Błachowski, K. Ruebenbauer, J. Żukrowski, S. M. Dubiel, L. M. Tran, M. Babij, and Z. Bukowski, J. Magn. Magn. Mater. **457**, 1 (2018).
[21] Anupam, V. K. Anand, P. L. Paulose, S. Ramakrishnan, C. Geibel, and Z. Hossain, Phys. Rev. B **85**, 144513 (2012).
[22] N. S. Sangeetha, Santanu Pakhira, D. H. Ryan, V. Smetana, A.-V. Mudring, and D. C. Johnston, Phys. Rev. Materials **4**, 084407 (2020).
[23] S. L. Bud'ko, V. G. Kogan, R. Prozorov, W. R. Meier, M. Xu, and P. C. Canfield, Phys. Rev. B **98**, 144520 (2018).
[24] Y.-B. Liu, Y. Liu, W.-H. Jiao, Z. Ren, and G.-H. Cao, Sci. China-Phys. Mech. Astron. **61**, 127405 (2018); M. A. Albedah, Z. M. Stadnik, O. Fedoryk, Y.-B. Liu, and G.-H. Cao, J. Magn. Magn. Mater. **503**, 166603 (2020).
[25] N. Kumar, S. Chi, Y. Chen, K. G. Rana, A. K. Nigam, A. Thamizhavel, W. Ratcliff II, S. K. Dhar, and J. W. Lynn, Phys. Rev. B **80**, 144524 (2009).
[26] S. R. Saha, N. P. Butch, K. Kirshenbaum, and J. Paglione, Phys. Rev. B **79**, 224519 (2009).
[27] I. Nowik, I. Felner, N. Ni, S. L. Bud'ko, and P. C. Canfield, J. Phys.: Condens. Matter **22**, 355701 (2010).
[28] I. Nowik and I. Felner, Physica C **469**, 485 (2009).
[29] I. Nowik, I. Felner, Z. Ren, Z. A. Xu, and G. H. Cao, J. Phys.: Conf. Ser. **217**, 012121 (2010).
[30] A. Błachowski, K. Ruebenbauer, J. Żukrowski, K. Rogacki, Z. Bukowski, and J. Karpinski, Phys. Rev. B **83**, 134410 (2011).
[31] J. Cieślak and S. M. Dubiel, Nucl. Instrum. Methods B **95**, 131 (1995).
[32] P. Bonville, F. Rullier-Albenque, D. Colson, and A. Forget, EPL **89**, 67008 (2010).
[33] K. Komędera, A. Pierzga, A. Błachowski, K. Ruebenbauer, A. Budziak, S. Katrych, and J. Karpinski, J. Alloys Compd. **717**, 350 (2017).
[34] J. M. Allred, K. M. Taddei, D. E. Bugaris, M. J. Krogstad, S. H. Lapidus, D. Y. Chung, H. Claus, M. G. Kanatzidis, D. E. Brown, J. Kang, R. M. Fernandes, I. Eremin, S. Rosenkranz, O. Chmaissem, and R. Osborn, Nature Phys. **12**, 493 (2016).
[35] W.-H. Jiao, Q. Tao, J.-K. Bao, Y.-L. Sun, Ch.-M. Feng, Z.-A. Xu, I. Nowik, I. Felner, and G.-H. Cao, EPL (Europhys. Lett.) **95**, 67007 (2011).
[36] M. A. Albedah, F. Nejadsattari, Z. M. Stadnik, Y. Liu, and G.-H. Cao, Phys. Rev. B **97**, 144426 (2018).
[37] M. A. Albedah, F. Nejadsattari, Z. M. Stadnik, Y. Liu, and G.-H. Cao, J. Phys.: Condens. Matter **30**, 155803 (2018).
[38] S. Ikeda, K. Yoshida, and H. Kobayashi, J. Phys. Soc. Jpn. **81**, 033703 (2012).
[39] S. M. Dubiel, J. Alloys Compd. **488**, 18 (2009).
[40] A. Błachowski, K. Ruebenbauer, J. Żukrowski, and Z. Bukowski, J. Alloys Compd. **582**, 167 (2014).
[41] K. Komędera, A. K. Jasek, A. Błachowski, K. Ruebenbauer, and A. Krztoń-Maziopa, J. Magn. Magn. Mater. **399**, 221 (2016).
[42] H. Raffius, E. Mörsen, B. D. Mosel, W. Müller-Warmuth, W. Jeitschko, L. Terbüchte, and T. Vomhof, J. Phys. Chem. Solids **54**, 135 (1993).
[43] Z. Guguchia, S. Bosma, S. Weyeneth, A. Shengelaya, R. Puzniak, Z. Bukowski, J. Karpinski, and H. Keller, Phys. Rev. B **84**, 144506 (2011).
[44] W. H. Jiao, I. Felner, I. Nowik, and G. H. Cao, J. Supercond. Nov. Magn. **25**, 441 (2012).
[45] Ch. Feng, Z. Ren, S. Xu, S. Jiang, Z. Xu, G. Cao, I. Nowik, I. Felner, K. Matsubayashi, and Y. Uwatoko, Phys. Rev. B **82**, 094426 (2010).

[46] L. Sun, J. Guo, G. Chen, X. Chen, X. Dong, W. Lu, Ch. Zhang, Z. Jiang, Y. Zou, S. Zhang, Y. Huang, Q. Wu, X. Dai, Y. Li, J. Liu, and Z. Zhao, Phys. Rev. B **82**, 134509 (2010).
[47] W. T. Jin, N. Qureshi, Z. Bukowski, Y. Xiao, S. Nandi, M. Babij, Z. Fu, Y. Su, and Th. Brückel, Phys. Rev. B 99, 014425 (2019).

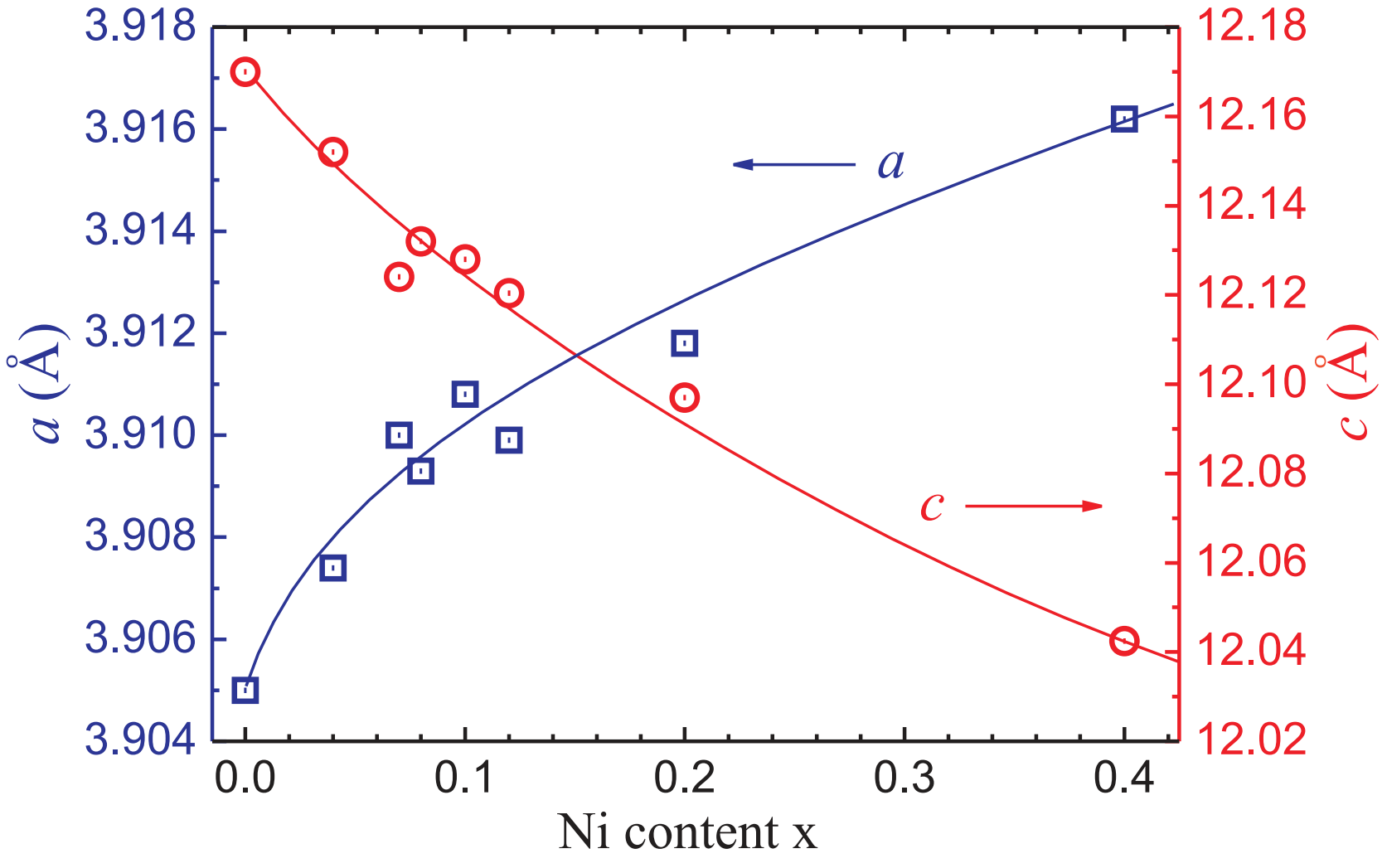

**Fig. 1** Lattice parameters of the tetragonal unit cell (space group *I*4/*mmm*) at room temperature for obtained $EuFe_{2-x}Ni_xAs_2$ samples.

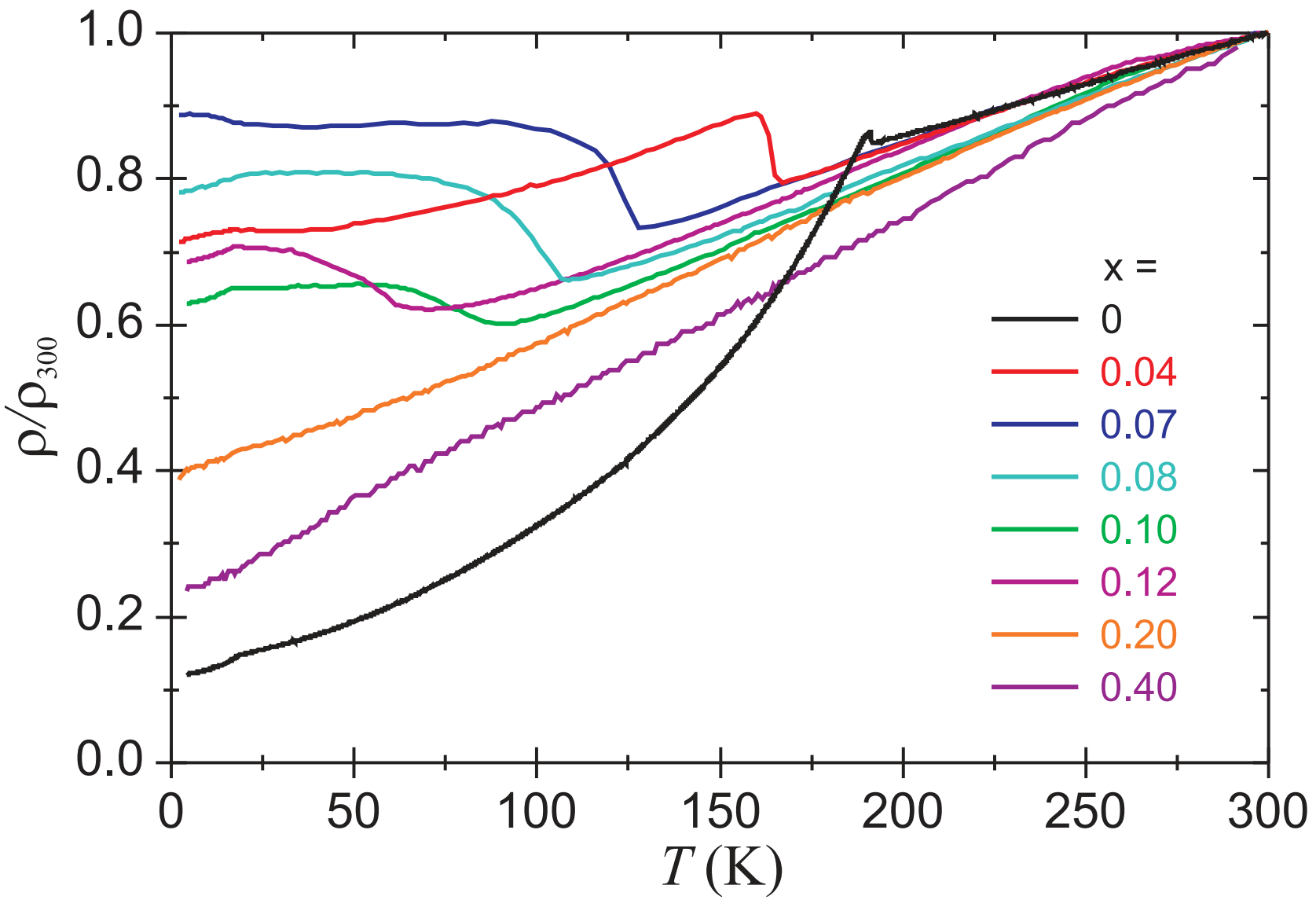

**Fig. 2** Temperature dependencies of the relative resistivity (normalized to the resistivity at 300 K) for $EuFe_{2-x}Ni_xAs_2$ samples. The current was applied perpendicular to the *c*-axis. The resistivity anomaly associated with the iron magnetism is clearly visible for all studied samples, except the highest Ni-substitution levels x = 0.20 and 0.40.

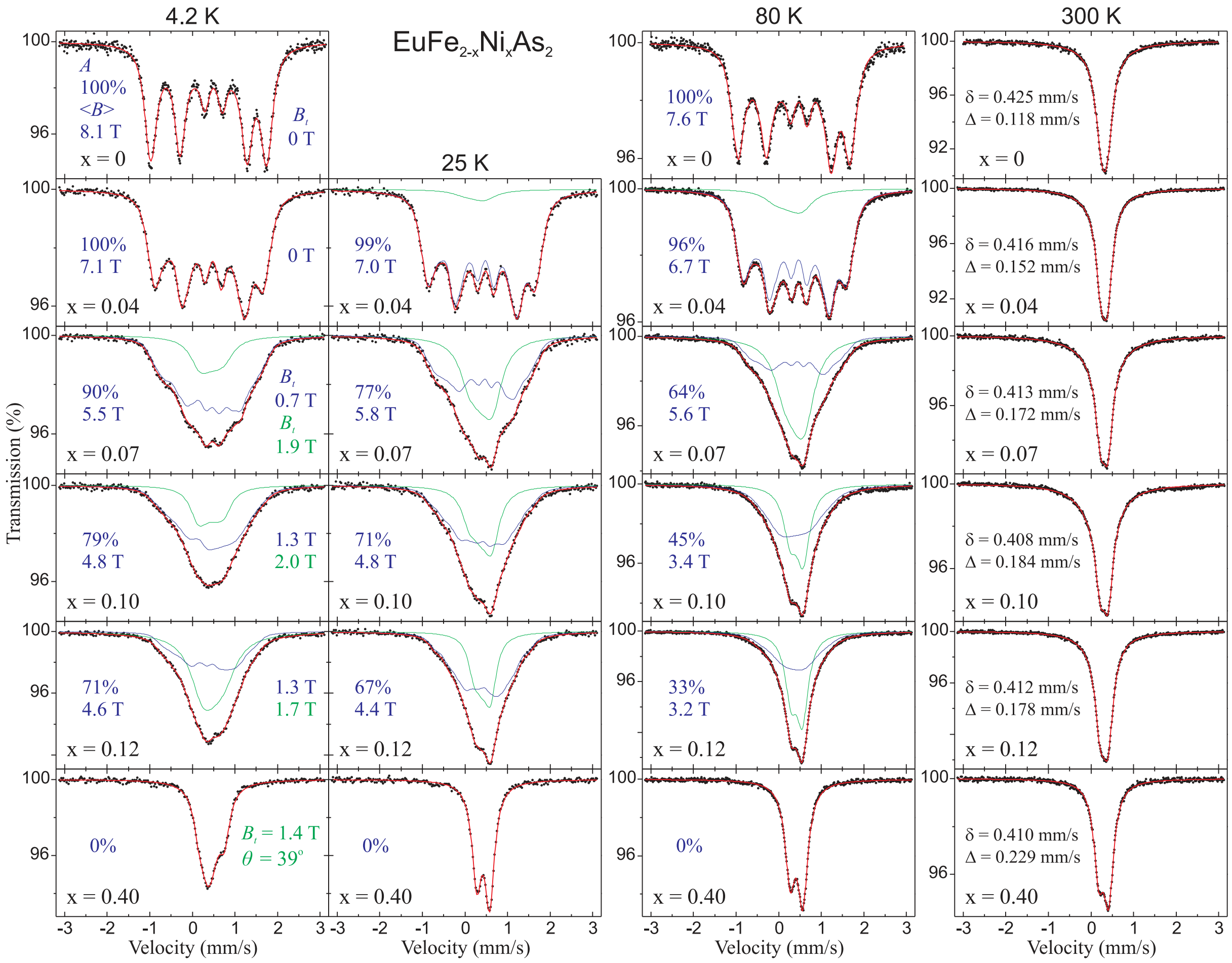

**Fig. 3** $^{57}$Fe Mössbauer spectra of $EuFe_{2-x}Ni_xAs_2$ measured at 4.2, 25, 80 and 300 K. (Inserted values in blue) The relative contribution of the SDW spectral component *A*, the average magnetic hyperfine field of SDW <*B*>, and the hyperfine transferred magnetic field due to the $Eu^{2+}$ ordering for iron in the SDW state. (Inserted values in green) The hyperfine transferred field due to the $Eu^{2+}$ ordering for iron in the "non-magnetic" state (the second spectral component), and the angle $\theta$ between the principal component of EFG at Fe nuclei (i.e. the crystallographic *c*-axis) and the $Eu^{2+}$ magnetic moment. The spectral center shift δ and the quadrupole splitting Δ at 300 K are shown. Errors for all values are of the order of unity for the last digit shown.

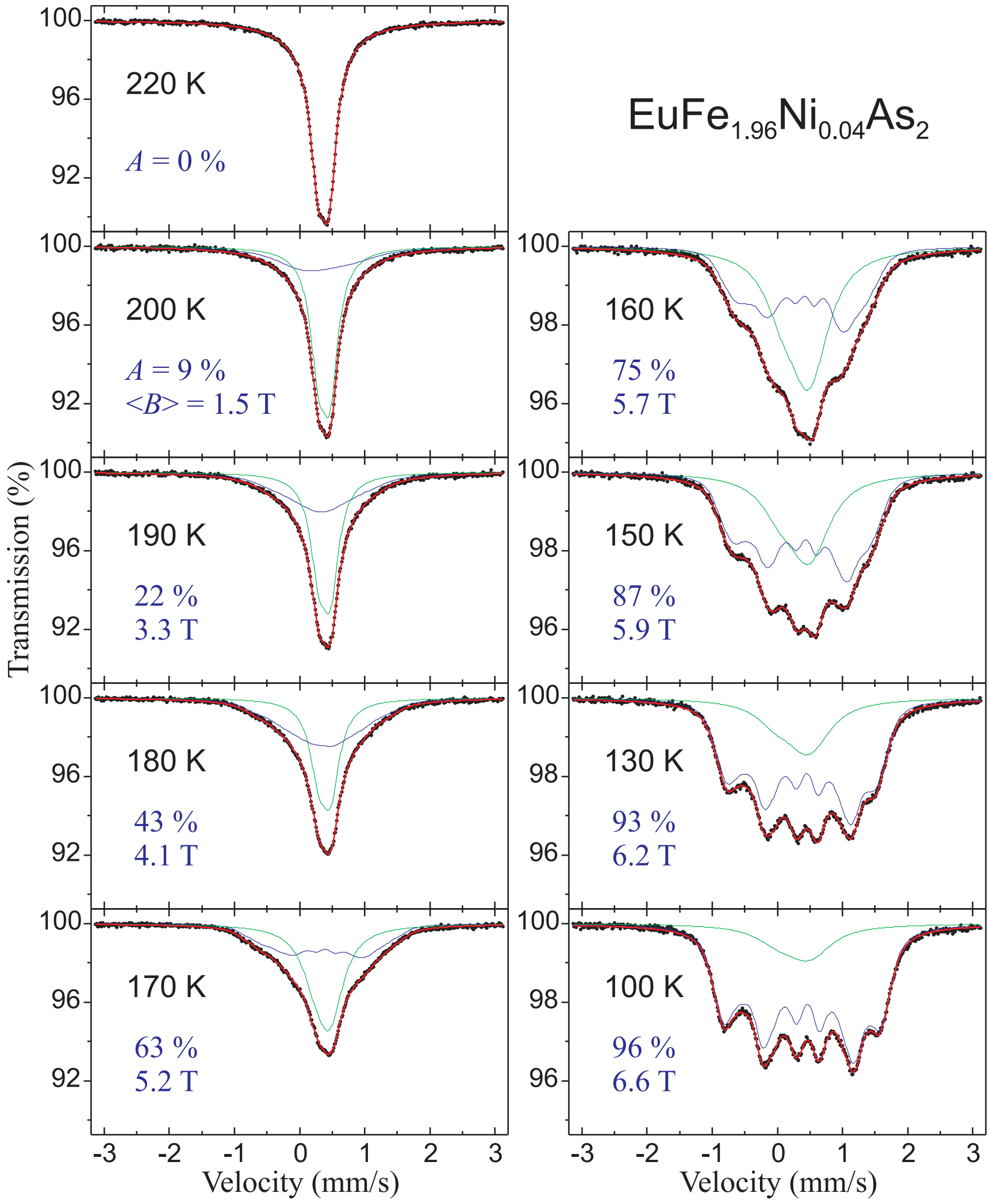

**Fig. 4** Selected $^{57}$Fe Mössbauer spectra of $EuFe_{1.96}Ni_{0.04}As_2$ versus temperature across the SDW order. (Inserted values) The contribution of the SDW spectral component *A*, and the average magnetic hyperfine field of SDW <*B*>. Spectra were measured at the same conditions during non-interrupted series and they are shown for comparison on the same scale, except that the left column scale is twice the scale of the right one. The influence of the magnetic SDW order on the shape of the spectra is clearly visible.

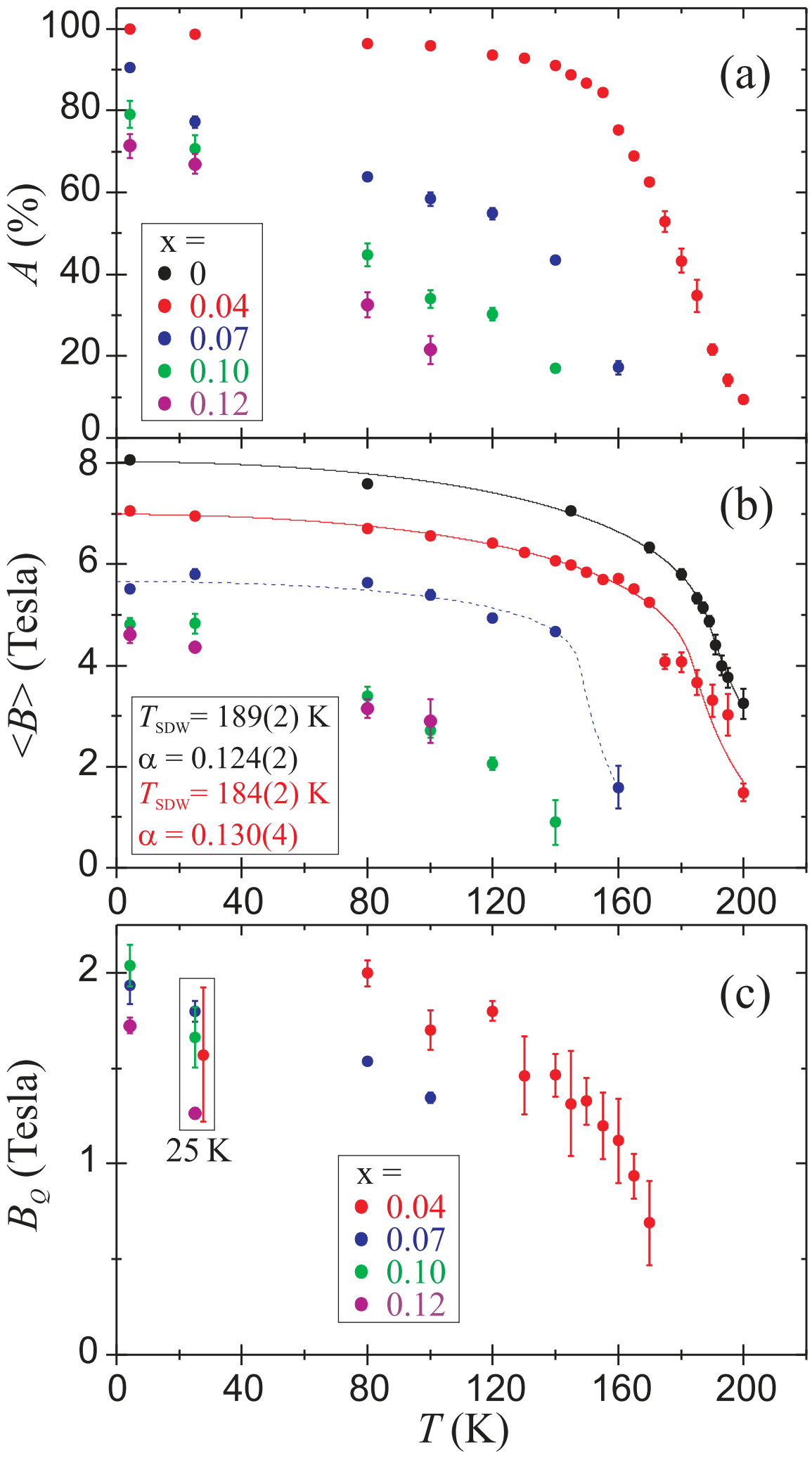

**Fig. 5** (a) Temperature dependence of the contribution of the SDW spectral component *A* for Ni-substituted compounds with x = 0.04, 0.07, 0.10, 0.12. (b) The average magnetic hyperfine field of SDW <*B*> versus temperature. The solid lines for x = 0 and 0.04 represent the best-fit to experimental data using the model described in Ref. [30]. The dashed line for x = 0.07 is for the view clarity only. (c) The low hyperfine field of the second spectral component $B_Q$ arising from the broadening of spectral lines.

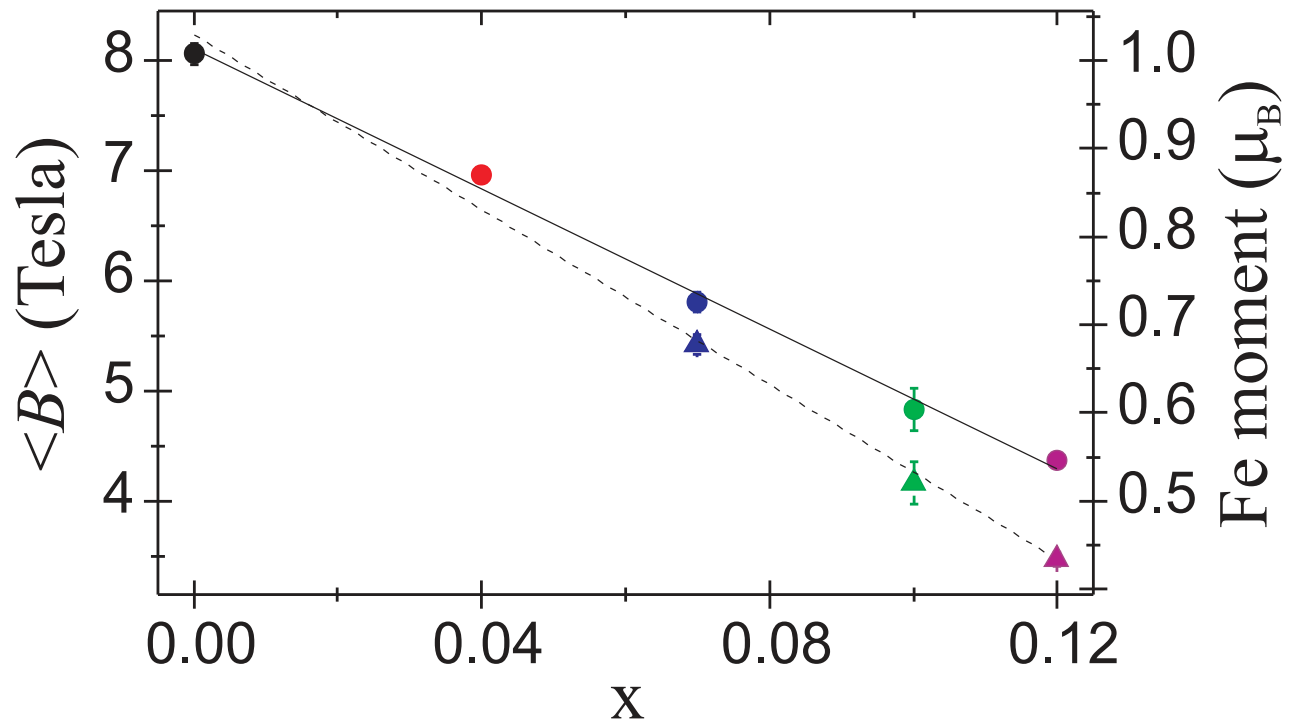

**Fig. 6** The average hyperfine field of SDW <*B*> at 4.2 K and corresponding itinerant magnetic moment versus Ni-substitution level x for $EuFe_{2-x}Ni_xAs_2$ system. Triangles represent the weighted average values of both quantities with taking into account the contribution of SDW phase.

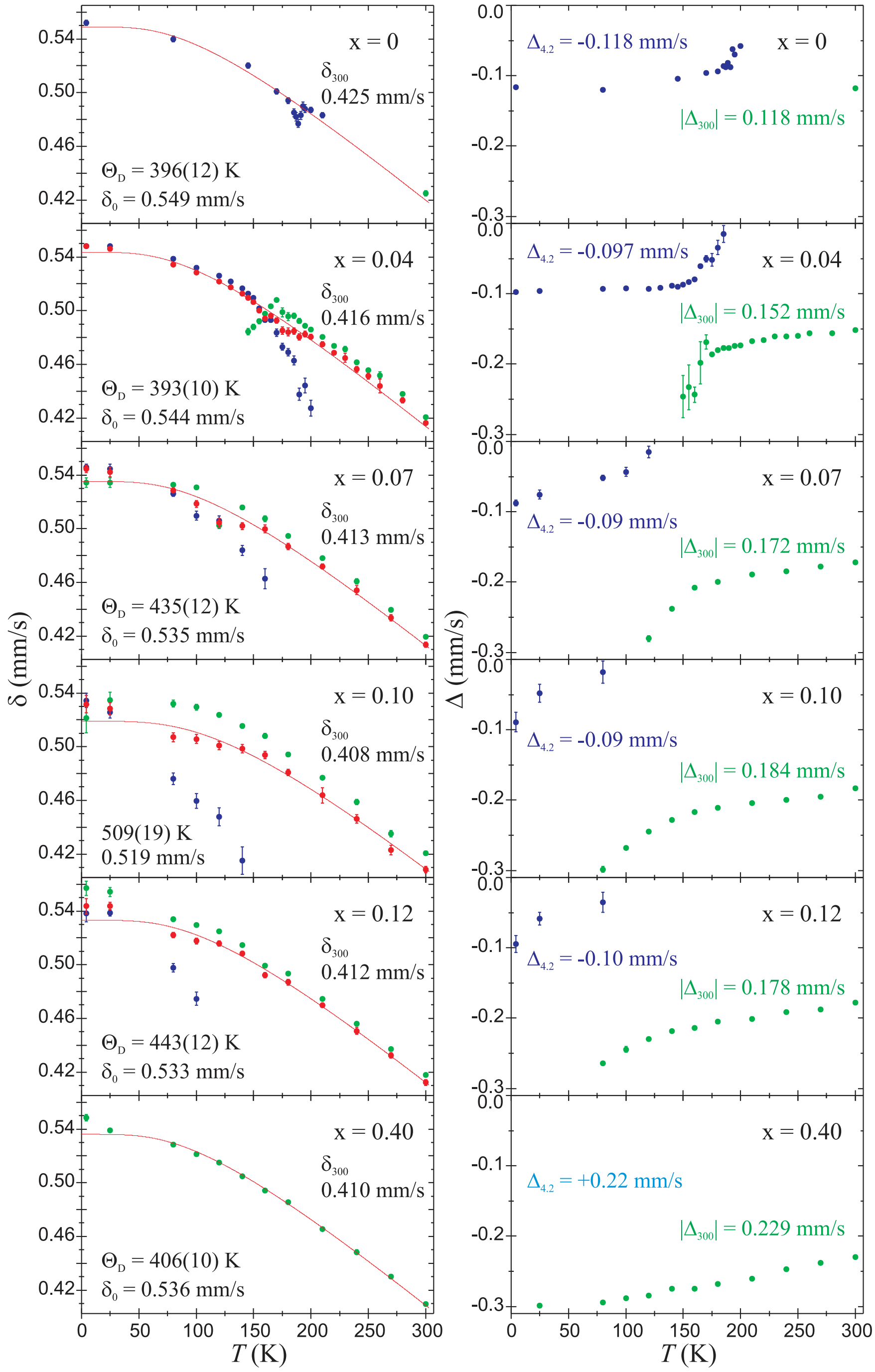

**Fig. 7** Temperature dependencies of the Mössbauer spectroscopy parameters: the spectral center shift δ relative to the shift of room temperature α-Fe, and the electric quadrupole splitting Δ. Values for the SDW spectral component and the second spectral component are shown in blue and in green, respectively. The weighted average spectral center shift <δ> is marked in red along with the solid lines representing the best-fit to experimental data using the Debye model for <δ>(*T*). The Debye temperature $\Theta_D$ and the spectral shift at the ground state $\delta_0$ are shown. The symbols $\delta_{300}$, $\Delta_{4.2}$ and $|\Delta_{300}|$ stand for measured values of the spectral shift, the quadrupole splitting, and the absolute value of the quadrupole splitting, at 4.2 K and 300 K, respectively. For x = 0.40, point of $\Delta_{4.2}$ with the positive sign determined in presence of the $Eu^{2+}$ transferred hyperfine field isn't shown. Note: for $^{57}$Fe the sign of the quadrupole coupling constant cannot be determined without the hyperfine field or applied external magnetic field.

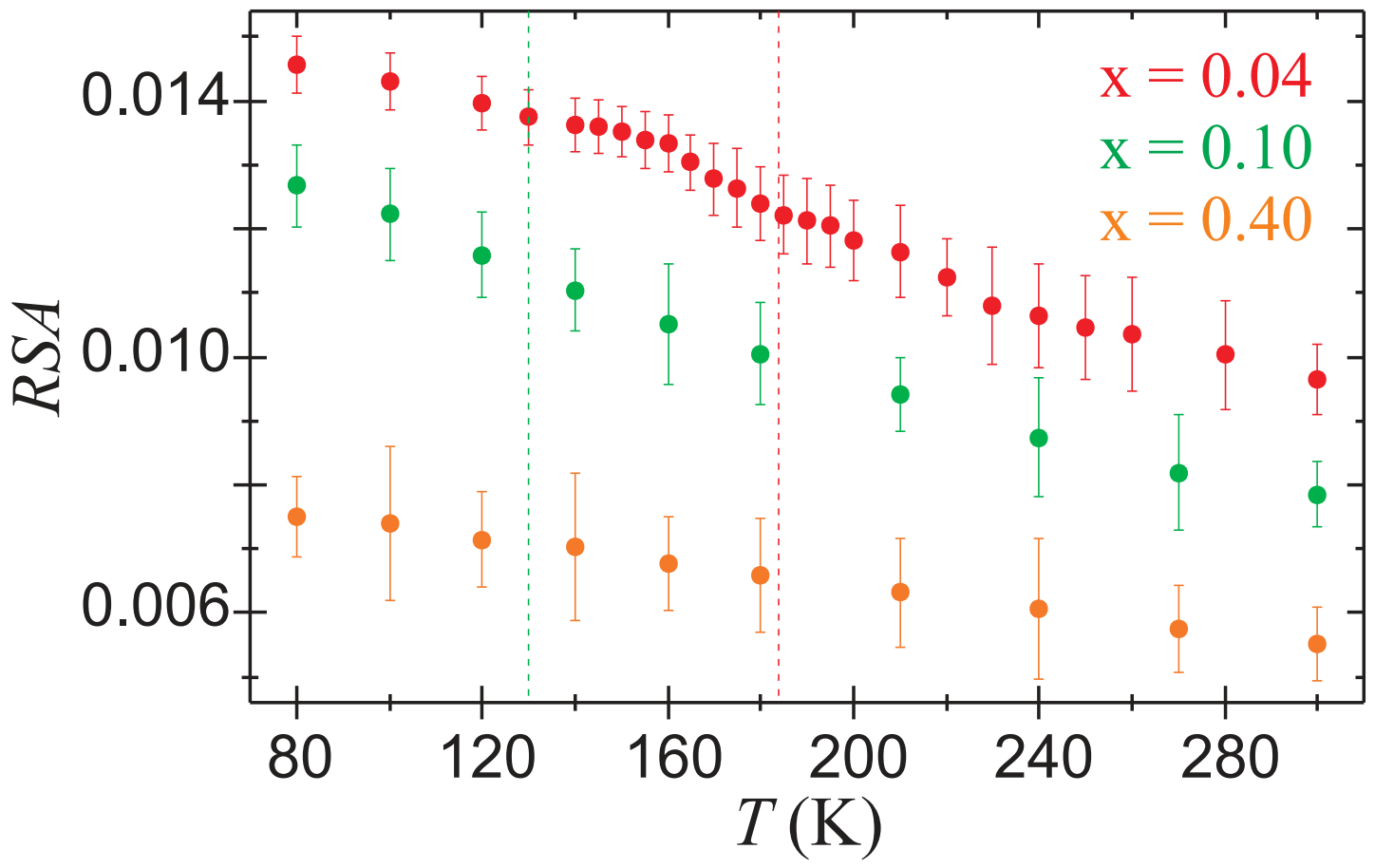

**Fig. 8** Temperature dependencies of the relative spectral area (*RSA*) for $EuFe_{2-x}Ni_xAs_2$ with x = 0.04, 0.10, 0.40. Vertical dashed lines at 184 K and 130 K denote the SDW order temperatures for x = 0.04 and x = 0.10, respectively. Each *RSA* set is derived from series of measurements performed at constant conditions (see text for details). Note: despite the identical weights of samples the sets of *RSA* differ due to difference in the resonant thickness ($^{57}Fe$ content) between particular absorbers.

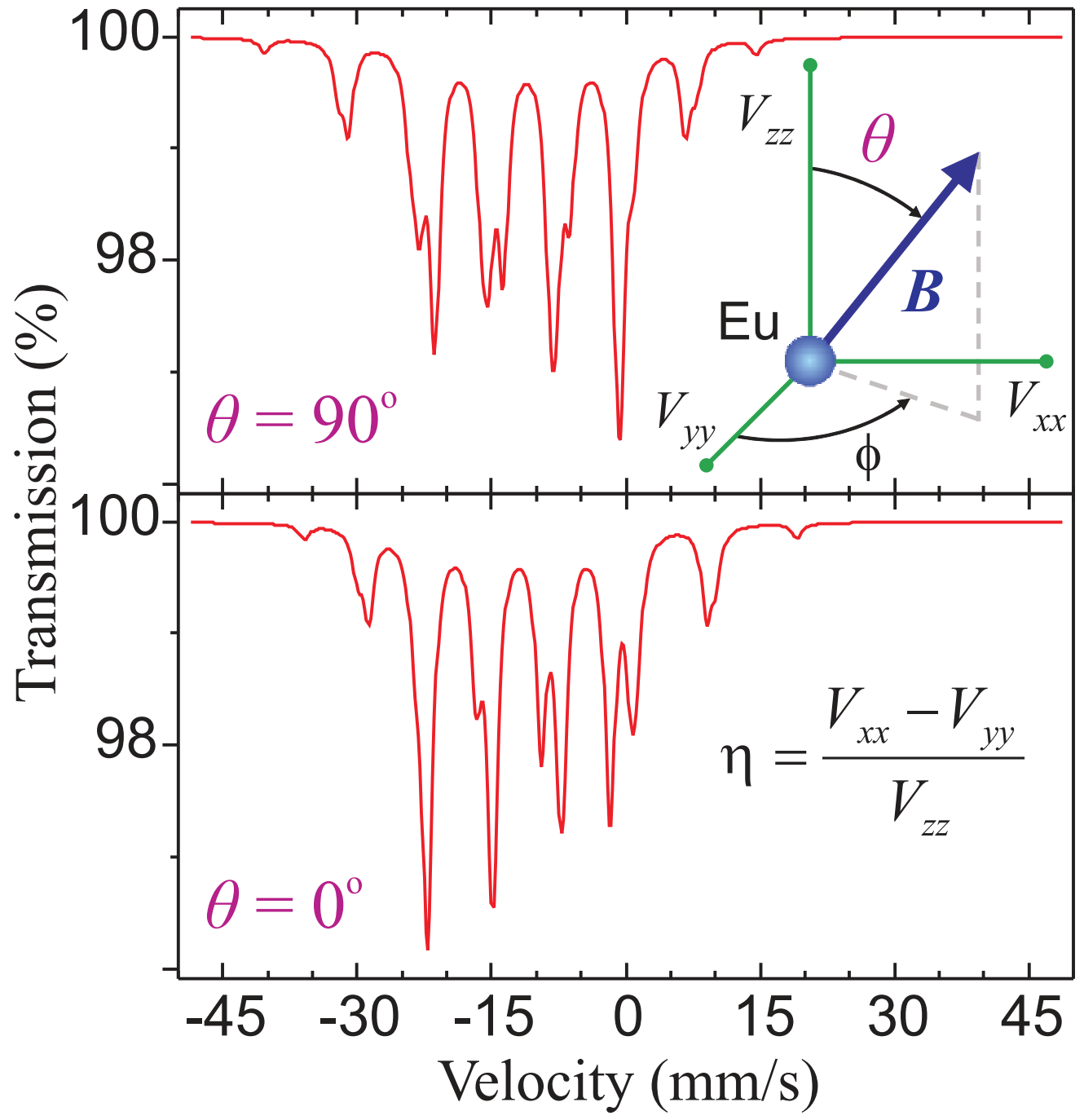

**Fig. 9** Simulated $^{151}Eu$ Mössbauer spectra for two extreme values of the $\theta$ angle, i.e. 90° and 0°. The other hyperfine parameters are constant and their values are close to those for $EuFe_2As_2$, i.e. $\delta$ = -11.4 mm/s, $V_{zz}$ = -0.49 ($10^{22}$V/m$^2$), and $B$ = 26.6 T. The spectral linewidths are equal to the natural width and the optimal effective absorber thickness was applied. Clearly visible change of the relative intensity of spectral lines illustrates the high sensitivity of $^{151}Eu$ Mössbauer spectroscopy to the phenomenon of spin reorientation. On the other hand, the $\phi$ angle is undetectable for the quadrupole asymmetry parameter $\eta = 0$, i.e. $V_{xx} = V_{yy}$, which is a case of tetragonal structure with $V_{zz}$ along *c*-axis of the unit cell.

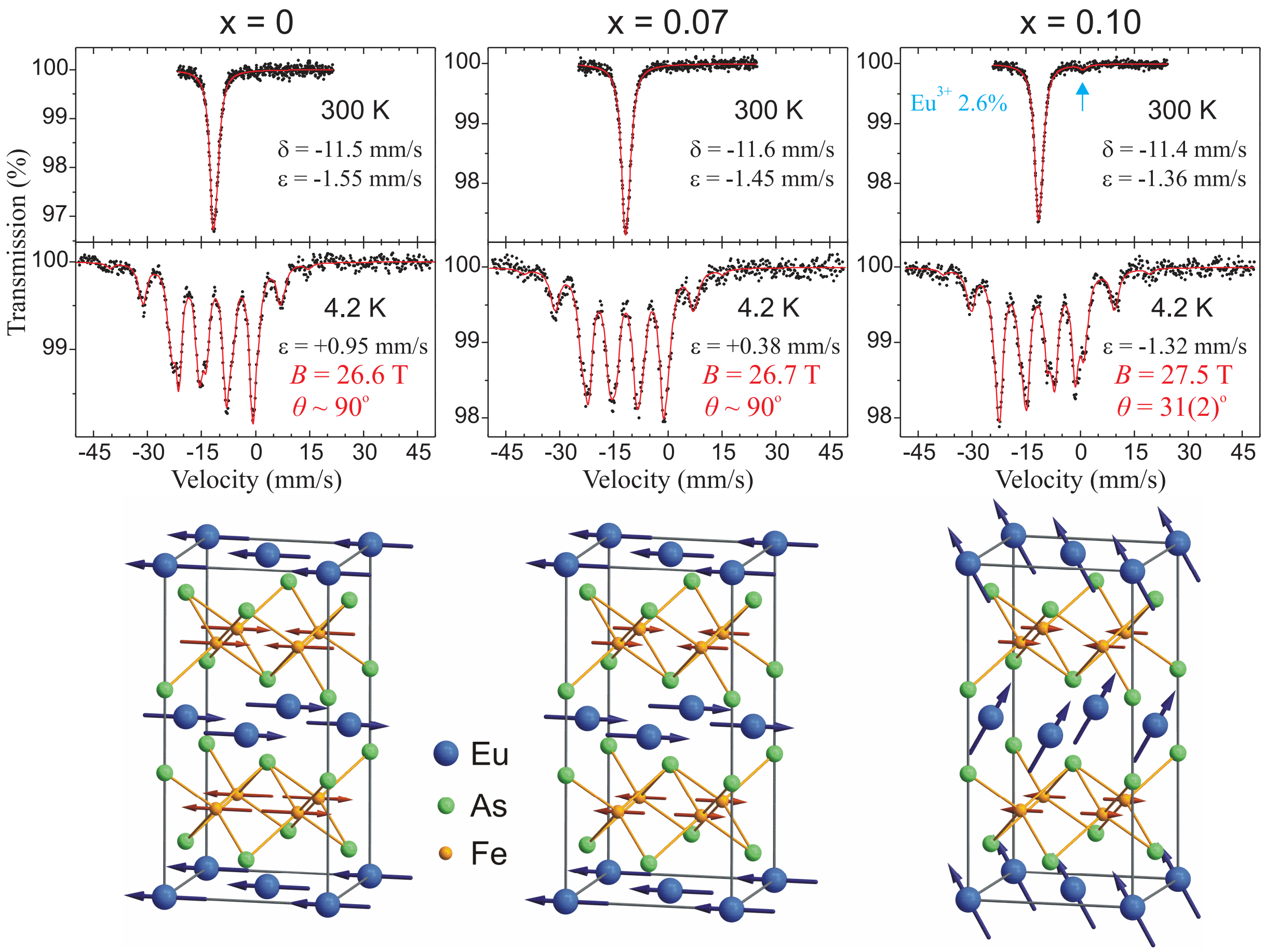

**Fig. 10** $^{151}$Eu Mössbauer spectra of $EuFe_{2-x}Ni_xAs_2$ with x = 0, 0.07, 0.10. The hyperfine magnetic field for $Eu^{2+}$ is denoted by $B$, while the symbol $\theta$ stands for the angle between the principal component of the EFG (*c*-axis of the unit cell) and the $Eu^{2+}$ hyperfine field (4*f*-Eu magnetic moment). The symbols δ and ε stand for the spectral center shift relative to room temperature $^{151}SmF_3$ source and the quadrupole shift, respectively. The arrow on spectrum of x = 0.10 at 300 K shows position of the $Eu^{3+}$ component with ε = +0.8(2) mm/s and $\Gamma_a$ = 1.6(6) mm/s. (Lower panel) Sketches of the corresponding magnetic structures of $EuFe_{2-x}Ni_xAs_2$ system at the ground state emerging from the $^{57}$Fe and $^{151}$Eu Mössbauer results at 4.2 K. The gray line outlines the orthorhombic unit cell.

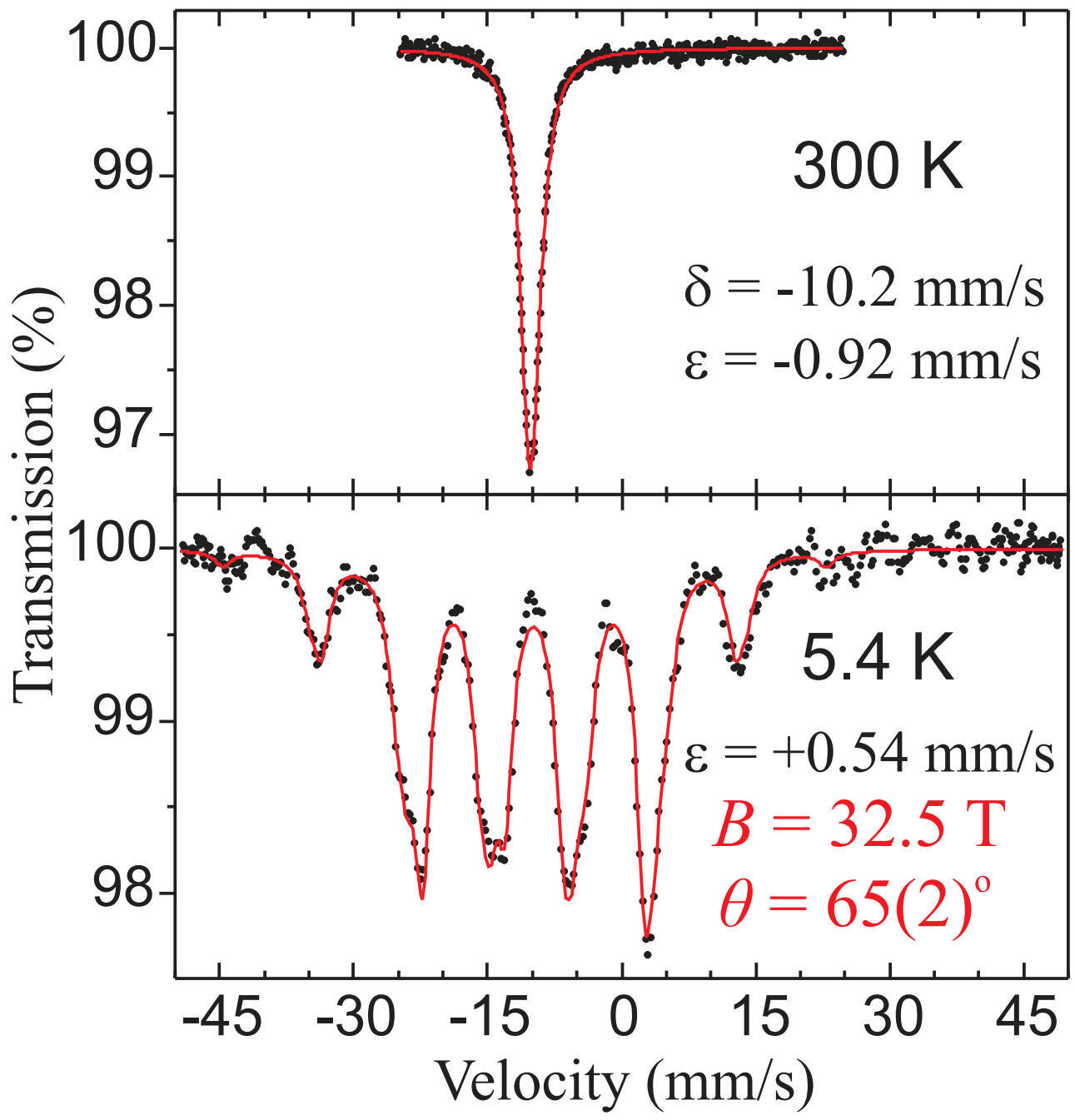

**Fig. 11** $^{151}$Eu Mössbauer spectra of $EuNi_2As_2$. The meaning of the symbols is the same as in Fig. 10. Description of the magnetic structure of this sample can be found in Ref. [47].